# Flexo-photovoltaic effect and above-bandgap photovoltage in halide perovskites

Zhiguo Wang[1, 6], Shengwen Shu[2, 6], Xiaoyong Wei[3], Renhong Liang[1], Shanming Ke[1], Longlong Shu[1] ✉, Gustau Catalan[4, 5] ✉

[1.] *School of Physics and Materials Science, Nanchang University, Nanchang 330031, People's Republic of China.*

[2.] *College of Electrical Engineering and Automation, Fuzhou University, Fuzhou 350108, People's Republic of China.*

[3.] *Electronic Materials Research Laboratory, Key Laboratory of the Ministry of Education and International Center for Dielectric Research, Xi'an Jiao Tong University, Xi'an 710049, People's Republic of China*

[4.] *Institucio Catalana de Recerca i Estudis Avançats (ICREA), Barcelona, Catalonia.*

[5.] *Institut Catala de Nanociencia i Nanotecnologia (ICN2), CSIC–BIST, Campus Universitat Autonoma de Barcelona, Barcelona, Catalonia.*

[6.] These authors contributed equally: Zhiguo Wang, Shengwen Shu.

✉e-mail: llshu@ncu.edu.cn; gustau.catalan@icn2.cat

ABSTRACT

Halide perovskites have outstanding photovoltaic properties which have been optimized through interfacial engineering. However, as these materials approach the limits imposed by the physics of semiconductor junctions, it is urgent to explore alternatives, such as the bulk photovoltaic effect, whose physical origin is different and not bound by the same limits [1-3]. In this context, we focus on the flexo-photovoltaic effect, a type of bulk photovoltaic effect that was recently observed in oxides under strain gradients [4]. We have measured the flexo-photovoltaic effect of $MAPbBr_3$ and $MAPbI_3$ crystals under bending and found it to be orders of magnitude larger than for SrTiO3, the benchmark flexo-photovoltaic oxide. For sufficiently large strain gradients, photovoltages bigger than the bandgap can be produced. Bulk photovoltaic effects are additive and, for $MAPbI_3$, the flexo-photovoltage exists on top of a native bulk photovoltage that is hysteretic, consistent with the electrically switchable macroscopic polarization of this material. The results suggest that harnessing the flexo-photovoltaic effect through strain gradient engineering can provide a functional leap forward for halide perovskites.

Generating a photovoltaic effect requires breaking space-inversion symmetry; otherwise photo-generated carriers do not have a preferential direction in which to flow (Neumann principle of symmetry). In conventional solar cells, spatial inversion symmetry is broken at interfaces, and electron-hole pairs generated at or near the interfaces move in opposite directions towards their most charge-affine side of the junction [5,6], thereby establishing a photocurrent. However, spatial inversion symmetry can also be intrinsically broken, without the need for interfaces, inside the bulk of some materials -piezoelectrics – which are defined by having non-centrosymmetric crystal structures. Electron-hole pairs photo-excited inside piezoelectrics have different probabilities of drifting in opposite directions and thus generate a shift current in response to illumination. This is known as the "bulk photovoltaic effect" [1-3].

In addition, the symmetry of any crystal, piezoelectric or otherwise, can also be broken by bending. This is the basis of flexoelectricity, a linear coupling between strain gradients and polarization first described in insulating dielectrics [7] but known to exist also in semiconductors [8]. Because bending deformation is inherently asymmetric (it creates tension on the convex side of the crystal and compression on the concave one), the so-called flexophotovoltaic effect [4] is theoretically possible in any material under bending. However, so far there has been no measurement of this effect in halide perovskites, a family of materials with large photovoltaic efficiency [6,9] which are known to be flexoelectric [10]. Here we report the flexophotovoltaic effect of halide perovskites $MAPbBr_3$ (hereafter MAPB) and $MAPbI_3$, and compare them to that of a reference oxide perovskite $SrTiO_3$ (STO), used as flexoelectric benchmark [11]. We find that strain gradients can induce large flexo-photovoltaic effects in halide perovskites and can even achieve above-bandgap photovoltages in these materials.

The halide-perovskite samples in this study are single crystals made by us (see Methods) and the STO crystals were commercially acquired (TOPVENDOR, Beijing, China). Identical Au electrodes were deposited on opposite sides of the crystals to make symmetric sandwich-capacitor structures, which were illuminated laterally so that the contributions from the two opposite electrode interfaces were of similar magnitude and opposite sign and hence mutually cancelled. Illumination was provided by LED light with 405 nm wavelength for MAPB and 365 nm for STO, which has a larger bandgap. The maximum light intensity was set to 1000 LUX, as calibrated at the sample location using a photometer. The same photometer, in combination with a polarizer placed between light source and sample, was used to verify that the light emitted by the LED was not polarized [see Supplementary video S1]. Rotating the polarizing filter also allowed characterizing the dependence of photovoltaic output on the polarization angle of incident light, which is important to determine the bulk nature of the photovoltaic effect [1,12]. The strain gradients required to induce a flexo-photoelectric response were generated by bending the crystals in single-clamp cantilever geometry. We used the standard continuum-mechanics equation to calculate the strain gradient from the vertical deflection [13]. The experimental scheme is depicted in Figure 1-A and the experimental details are provided in the Methods section.

In the absence of bending, the close-circuit photocurrent and open-circuit photovoltage of STO and MAPB are found to be negligible, as expected for cubic centrosymmetric materials, and sizeable photovoltage and photocurrents appear only when the crystals are bent (Figure 1). The use of single crystals with a well-defined bending geometry allows quantifying the flexophotovoltaic effect and compare it between different materials. We define the flexophotovoltage coefficient as $\Phi_V \equiv \frac{\partial V_{OC}}{\partial G}$, where $V_{OC}$ is the open-circuit photovoltage at 1000 LUX, measured in volts, and $G$ is the strain gradient measured in $m^{-1}$. For STO, we find $\Phi_V = -3\times10^{-3}$ $V/m^{-1}$ from linear regression of the photovoltage as a function of strain gradient in figure 1-D. In contrast, the flexo-photovoltaic coefficient of MAPB, measured in identical conditions as the STO crystal, is much larger: $\Phi_V$ = -1.3 $V/m^{-1}$ (Figure 1-H).

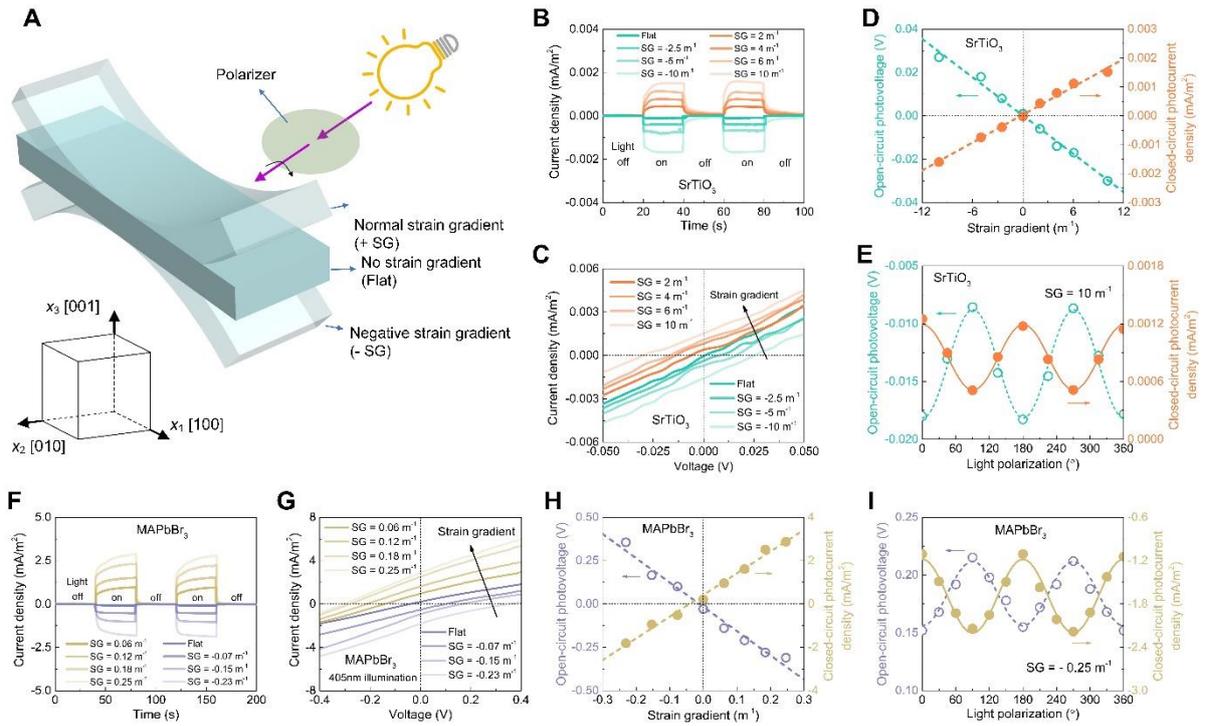

**Figure 1. Schematic configuration of flexo-photovoltaic experiment and results for STO and MAPB.** (A) Single crystals with identical top and bottom electrodes are clamped in the thickness direction [001] and bent vertically (see Supplementary materials figure S1). The lower left shows the relation between coordinate axis and crystal direction. The crystals were bent around the [010] axis and illuminated from the side parallel to the bending axis. A polarizing filter was introduced after the measurements in order to verify the bulk nature of the photoresponse. All measurements were performed under 1000 LUX of illuminance. (B and F) photocurrent density as a function of applied strain gradient. (C and G) current as a function of voltage under bending and illumination. (D and H) open-circuit photovoltage (voltage for which there is no current) and closed-circuit photocurrent (current at 0 V) as a function of strain gradient, extracted from (C and G). Linear fits to the data yield the flexophotovoltaic and flexophotocurrent coefficients, respectively. (E and I) photovoltage and photocurrent as a function of the polarization of the incident light with respect to the [001] direction, which is perpendicular to the electrodes and parallel to the flexoelectric polarization. The observed pi-periodic response is consistent with a bulk photovoltaic effect.

These results demonstrate that bending induces a photovoltaic effect in MAPB. Although photovoltaic contributions from crystal-electrode interfaces are self-cancelling when the crystal is flat (they are of equal magnitude and opposite sign), their equivalence is broken when the crystal is bent because one interface is under compression while the other is under tension, and this can yield a response that is flexoelectric-like but of interfacial rather than bulk origin [10,14,15]. We therefore need to clarify whether the observed bending-induced photovoltage is caused by a bulk flexo-photovoltaic effect or a by an interfacial change in the deformation potentials. Figures 1-E and 1-I show that the photovoltage and photocurrent change as a function of the polarization angle of incident linearly-polarized light. These variations are not due to changes in the intensity of the incident light (see Methods and Supplementary video S1). The observed 90-degree dependence of photovoltage and photocurrent on light-polarization angle is a trademark signature of the bulk photovoltaic effect [1,12] and the flexo-photovoltaic effect [4].

To explore the generality of the results, we have extended this investigation to another halide perovskite, the archetype methil-ammonium lead iodide (MAPbI$_3$, henceforth MAPI). Compared to MAPB, this material has a smaller bandgap and a crystal structure that might be ferroelectric at room temperature, although there is no unanimity about this aspect [16-18]. The MAPI single crystals were made and measured following the same procedure as for MAPB, and the results are shown in Figure 2 A-B. Like in MAPB, we observe a clear flexo-photovoltaic effect, in the sense that bending the crystal modifies the photovoltaic output of the crystal. However, unlike MAPB, MAPI shows a non-zero photovoltaic output even in the initial flat state. This implies that the MAPI crystal is already polar to start with, with a native bulk photovoltaic effect that is then further modulated by bending. The bulk nature of both these photovoltages is supported by the 90-degree sinusoidal dependence of the photovoltage on the light polarization, shown in the supplementary materials Figure S2.

To gain further insight into the native bulk photovoltage of MAPI, we measured the open-circuit photovoltage as a function of poling field. The results show a ferroelectric-like hysteretic response (Figures 2 C-D). These photovoltages and photocurrents have been measured after, not during, the application of poling voltage, and the hysteresis loops are therefore robust against the so-called "banana loop" artifact that affects conventional hysteresis measurements of polarization as a function of field in non-insulating materials [19]. The results indicate that MAPI crystals have two polar states separated by an energy barrier and switching the macroscopic polarization reverses the open-circuit photovoltage. From a functional point of view, therefore, MAPI is ferroelectric-like, and the photovoltage dependence on its poling history means that, in addition to strain gradients, external voltages can be used modify its photovoltaic output. On the other hand, these measurements can prove macroscopic polarization but cannot determine whether ferroelectricity exists at the crystallographic unit-cell level. The distinction is important because macroscopic polarization, induced by chemical or micro-structural gradients, can exist in cristallographically non-polar materials [20,21], and in the case of MAPI it is known that macroscopic polarization can be induced induce via ionic separation [22].

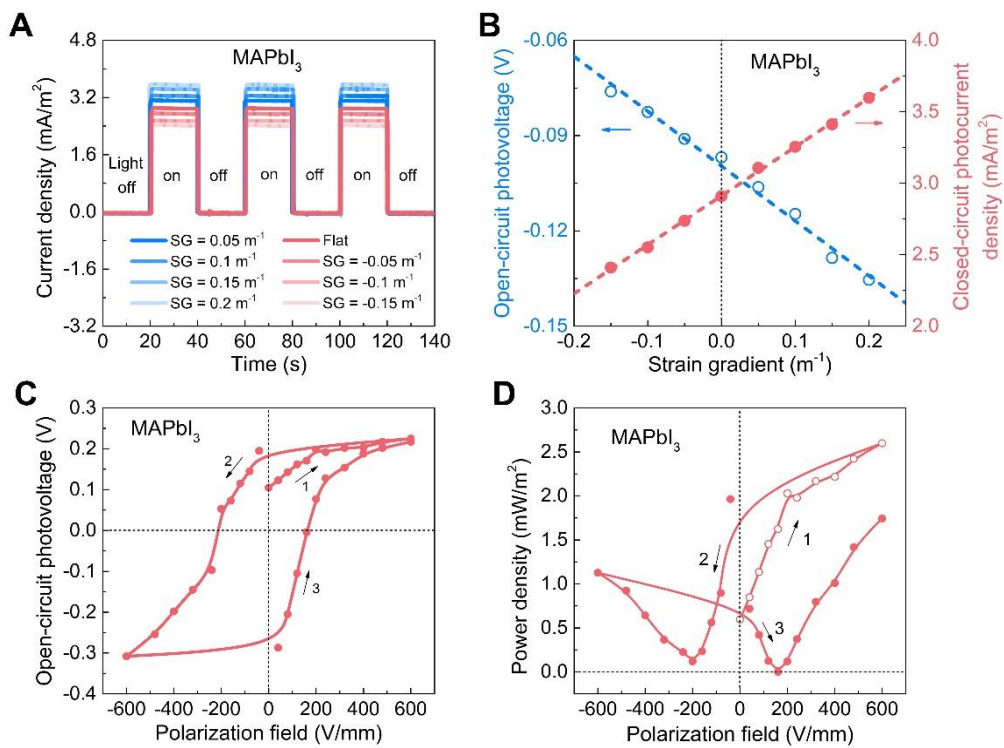

**Figure 2**. Photovoltaic output of MAPI as a function of strain gradient (A, B) and poling field (C, D). The results show a hysterestic bulk photovoltaic effect that can be modified both by electric fields and by strain gradients.

Contrary to the photovoltage of a semiconductor junction, the open-circuite voltage generated by the bulk photovoltaic effect can be bigger than the bandgap [1,23,24] and, since the flexophotovoltaic effect in of halide perovskites is a bulk photovoltaic effect, it is natural to wonder whether strain gradients can induce above-bandgap photovoltages. In our bulk crystals, the maximum curvature that could be generated without breaking was ≤ 0.25 m$^{-1}$, and this was insufficient to induce an above-bandgap photovoltage. On the other hand, strain gradients scale in inverse proportion to relaxation length and it is therefore possible to induce larger strain gradients by working at smaller size scales [25,26]. In particular, pressing a material with the sharp tip of an atomic force microscope is known to induce very large local strain gradients and flexoelectricity [27, 4]. We have therefore made a smaller (micron-thick) crystal and used an atomic force microscope with a conductive tip to induce a large vertical strain gradient and collect the photovoltaic output as a function of the vertical force applied to the tip (Figure 3).

The open-circuit photovoltage was found to increase in direct proportion to the indentation force and therefore the strain gradient, exceeding the 2.3eV bandgap around 1 µN and reaching ~4 eV for the applied forces of the order of 10 µN (Figures 3-C and 3-D). Even larger photovoltages of up to ~6V, which is more than twice the bandgap, could be achieved at lower indentation forces by working with a sharper tip (see supplementary materials S6). Since a larger-than-bandgap photovoltage cannot be generated in a semiconductor interface and our samples are single crystals free from internal grain boundaries that might act as tandem junctions [28], the results further support the bulk nature of the flexo-photovoltaic effect of MAPB, also verified by the 90-degree sinusoidal dependence of the photovoltage on light-polarization angle (inset of Figure 3-D).

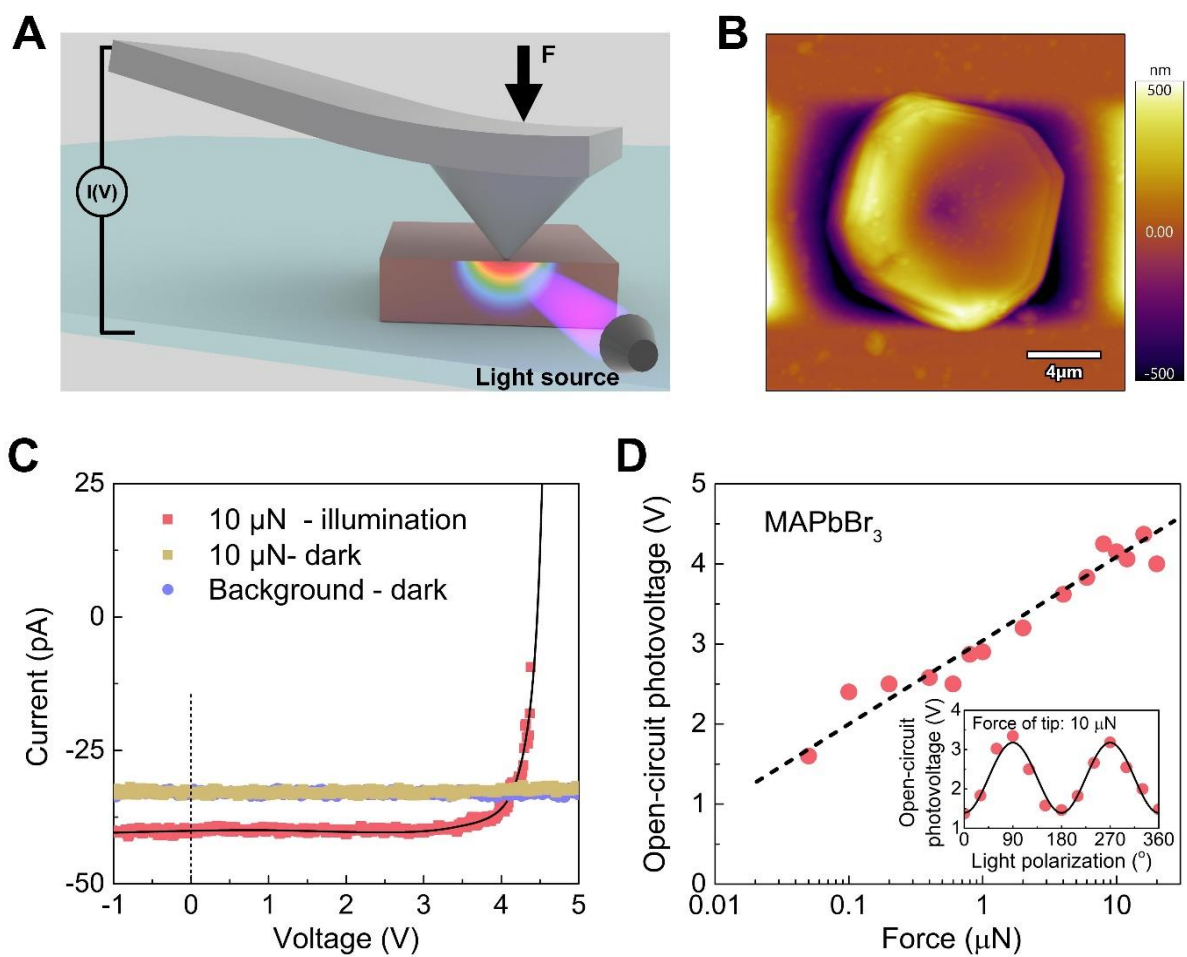

**Figure 3**: Above-bandgap photovoltage induced by flexoelectricity. (A) scheme of the experiment, whereby a large local curvature is induced by pushing with the tip of an atomic force microscope, and the current vs voltage is measured under illumination for various indentation forces. (B) micro-crystal useed for the measurement. The thickness of the crystal is ~1 μm and the bottom electrode is ITO. We observed no mechanical damage to the sample even after applying a maximal indentation force of 20 μN (see comparison in supplementary materials **Figure S2.** (C) current vs voltage measured under an indentation force of 10 μN, showing a ~4V open-circuit photovoltage (D) open-circuit photovoltage as a function of indentation force, showing the correlation between the two and (inset) photovoltage as a function of linear light-polarization angle, showing the same sinusoidal dependence as in the bulk crystals.

Although strain gradient engineering is a less mature field than strain engineering [29], strategies are emerging. Flexible free-standing films [30] can be deposited onto pre-tensed polymer substrates to induce wrinkles [31], or self-roll into cylinders with radial strain gradients [32]. Even in conventional flat-film architectures, vertical gradients can be induced through strain relaxation [33] or compositional gradients [34], while in polycrystalline films and ceramics strong internal strain gradients exist at grain boundaries [35], meaning that the flexo-photovoltaic effect may already be playing a role in the photovoltaic output of polycrystalline perovskites. At any rate, the strength of the flexophotovoltaic effect provides a clear motivation for exploring the limits of strain gradient engineering in halide perovskites.

**Acknowledgments:**

This work was supported by the Natural Science Foundation of China (Nos. 12174174 and 51962020) and the Natural Science Foundation of Jiangxi Province (Nos. 20212ACB214011 and 20202ZDB01006). GC acknowledges financial support from Project PID2019-108573GB-C21 funded by MCIN/AEI/10.13039/501100011033 and a Severo Ochoa Grant CEX2021-001214-S. He also thanks Prof. Dennis Meier for hosting GC as visiting professor at NTNU (Trondheim) during part of the time invested in this research. L.S thanks the support from Nanchang University. We thank Prof. Massimiliano Stengel for stimulating discussions.

**Methods**

**Sample information:**

STO (100) single crystals were commercially acquired from TOPVENDOR (www. topvendor.com.cn), with dimensions of 5 mm × 15 mm × 0.1 mm (width × length × thickness). The dimensions of electrodes are 5 mm × 4 mm (width × length).

MAPB single crystals for bending were prepared using these raw materials: methylammonium bromide ($CH_3NH_3Br$, 99.5%, Advanced Election Technology) or iodide ($CH_3NH_3I$, 99.5%, Advanced Election Technology) lead bromide ($PbBr_2$, 99.9%, Advanced Election Technology), N,N-dimethylformamide (DMF, 99%, Energy Chemical). MAPI single crystals for bending were prepared using these raw materials: methylammonium iodine ($CH_3NH_3I$, 99.5%, Advanced Election Technology), Lead iodine ($PbI_2$, 99.5%, Advanced Election Technology), γ-butyrolactone (GBL, ≥ 99%, Energy Chemical). MAPB and MAPI single crystals were grown using the inverse temperature crystallization method [36]. MAPB and MAPI single crystals were grown from its saturated solution in DMF and GBL (the concentration is 1 mol/L), respectively. We used 2000 mesh sandpaper to sand down the crystals to a lower thickness, followed by polishing the top and bottom surfaces with finer sandpaper (first 5000 mesh, then 10000 mesh) to achieve a mirror-smooth surface. The final dimensions of the MAPB single crystals were 5 mm × 15 mm × 1 mm, and MAPI single crystals are 5 mm × 15 mm × 0.9 mm for flexo-photovoltaic. The dimensions of electrodes are 5 mm × 4 mm (width × length).

MAPB micro-crystals for AFM experiment: The MAPB solution (1 mol/L) was spin-coated at 2000 rpm/min for 30s on the glass with indium tin oxide (ITO) electrode. Heating treatment at 70 °C for 30 min was carried out after the spin-coating process. The thickness of the micro MAPB single crystals is ~1 μm (Figure S3).

**Illumination**

The photovoltaic response was induced using a non-polarized LED light source (Model: HY-UV0003). The illuminance on the sample was measured using a photometer (Model: DT1330A, LIHUAJIN). The non-polarized nature of the light was verified by measuring the intensity of the incident light as a function of polarizer angle; the intensity variations as a function of polarizer rotation were smaller than 0.6% (see supplementary video S1).

**Strain Gradients**

The rectangular-beam-shaped crystals were clamped at one end and pushed at the opposite free end to induce single-clamp cantilever bending. The free end was pushed either from above or from below to induce positive or negative strain gradients respectively. In single-clamp cantilever geometry, the strain gradient is calculated from the vertical deflection of the free end using this equation: [13]

$$\frac{\partial \varepsilon_{11}}{\partial x_3} = \frac{3w(L)}{L^2}\left(1 - \frac{x}{L}\right) \quad (1)$$

where *w*(*L*) is the vertical deflection delivered by the piezoelectric actuator at the end of the cantilever, *L* is the length of the cantilever and *x* is the horizontal position from the center of the electrode to the fixed end, respectively.

**Atomic Force Microscopy (AFM)**

The electrical-transport characterizations of MAPB single crystal under a tip force was measured using a Cypher ES (Asylum Research) atomic force microscope (AFM) with ORCA mode (conductive AFM). A diamond-coated conductive probe (Model: CDT-NCLR) with a diameter of ~150 nm and force constants of 72 N m$^{-1}$ was used to apply pressure to the MAPbBr$_3$ single crystal, and the current-voltage (I-V) curve of the crystal was measured at the same time. An electrical bias was applied through the substrate, which was swiped at a ramping rate ~1 V s$^{-1}$. The background noise is lower than ± 50 pA in the 20 nA measurement range. (1) The AFM tip was placed in contact with the surface of the crystal with a preset force. The applied force was increased by further deflecting the AFM tip cantilever. (2) After the applied force reached a preset value, the compressive force was maintained constant by the feedback amplifier circuit of the controller, and then the electrical measurements were initiated. In this stage, a sweeping bias was applied between the substrate and the tip. (3) After the I–V measurement had been performed, a new compressive force was applied by setting a new target set point value and the corresponding electrical measurement was performed. In addition, we performed measurements of current and voltage vs time for given set-points. The results, shown in the supplementary materials figure S6, show that the steady-state values do not differ significantly from those measured in the I-V cycles.

The AFM tips are calibrated by the thermal noise method. We can precisely obtain the inverse optical lever sensitivity (InvOLS) and the spring constant of the tips. The precise force is calculated by multiplying the spring constant, InvOLS and cantilever deflection (force = spring constant × InvOLS × (preset deflection – initial deflection)). All measurements were performed in air at room temperature.

Supplementary Materials for

Flexo-photovoltaic effect and above-bandgap photovoltage in halide perovskites

Zhiguo Wang[1, 6], Shengwen Shu[2, 6], Xiaoyong Wei[3], Renhong Liang[1], Shanming Ke[1], Longlong Shu[1] ✉, Gustau Catalan[4, 5] ✉

[1.] School of Physics and Materials Science, Nanchang University, Nanchang, People's Republic of China.

[2.] College of Electrical Engineering and Automation, Fuzhou University, Fuzhou 350108, People's Republic of China.

[3.] Electronic Materials Research Laboratory, Key Laboratory of the Ministry of Education and International Center for Dielectric Research, Xi'an Jiao Tong University, Xi'an 710049, People's Republic of China

[4.] Institucio Catalana de Recerca i Estudis Avançats (ICREA), Barcelona, Catalonia.

[5.] Institut Catala de Nanociencia i Nanotecnologia (ICN2), CSIC–BIST, Campus Universitat Autonoma de Barcelona, Barcelona, Catalonia.

[6.] These authors contributed equally: Zhiguo Wang, Shengwen Shu.

✉ e-mail: llshu@ncu.edu.cn; gustau.catalan@icn2.cat

## S1. Intensity of incident light *vs* polarizer angle

Polarization.mp4

The illuminance on the sample is nearly independent of the polarizer angle (the maximum variation is <0.6%), confirming that the light emitted by the LED is not linearly polarized. Variations of photovoltage as a function of light-polarization are therefore a response of the material and not an illuminance artifact.

## S2. Cantilever experimental setup

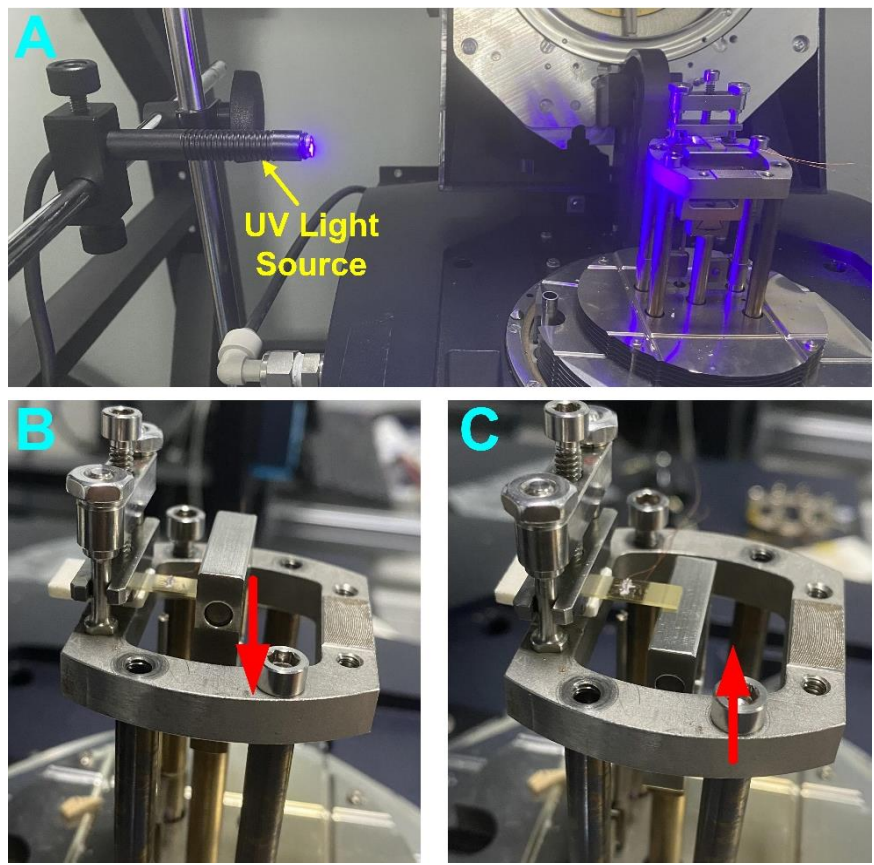

**Figure S1.** (A) The flexo-photovoltaic effect was characterized in single-clamp cantilever-bending geometry with lateral illumination. (B) Positive strain gradients are induced by locating the displacement module above the sample and setting a downward displacement. (C) Negative strain gradients are induced by locating the displacement module below the sample and setting an upward displacement. By swapping between (B) and (C), the sign of the strain gradient is changed whilst ensuring that the position of the sample relative to the light source is not altered.

**S3. Light-polarization dependence of photovoltaic output for MAPbI₃ single crystal**

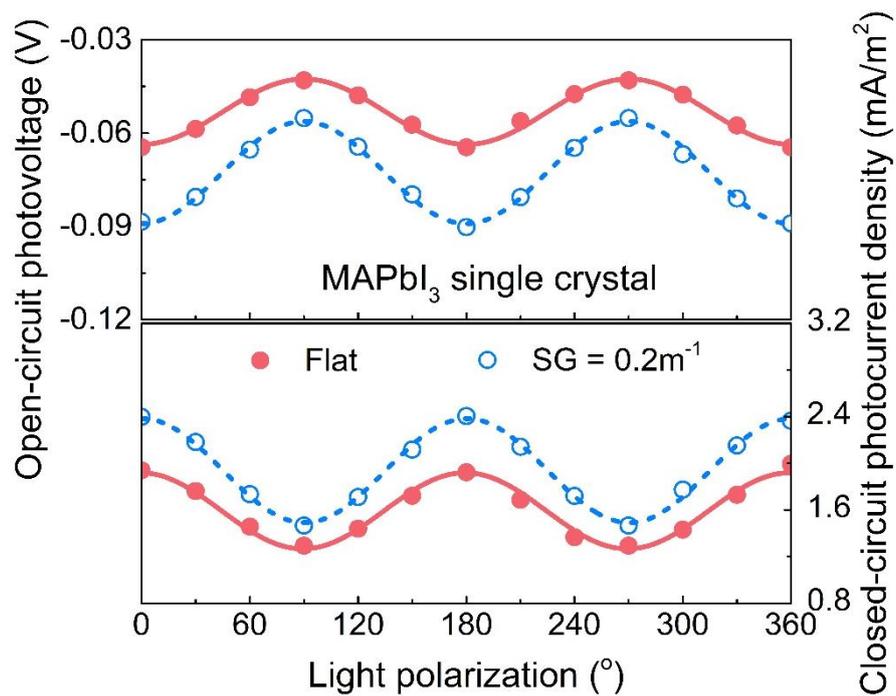

**Figure S2.** 90-degree sinusoidal modulation of photovoltaic output for MAPbI₃ single crystal as a function of light polarization for the crystal in flat (continuous line) and bent (dashed line) states. The results indicate that, already in the flat state, the photovoltaic output is polarization-dependent, consistent with the existence of a native macroscopic polarization.

## S4. The effect of the AFM tip on the surface

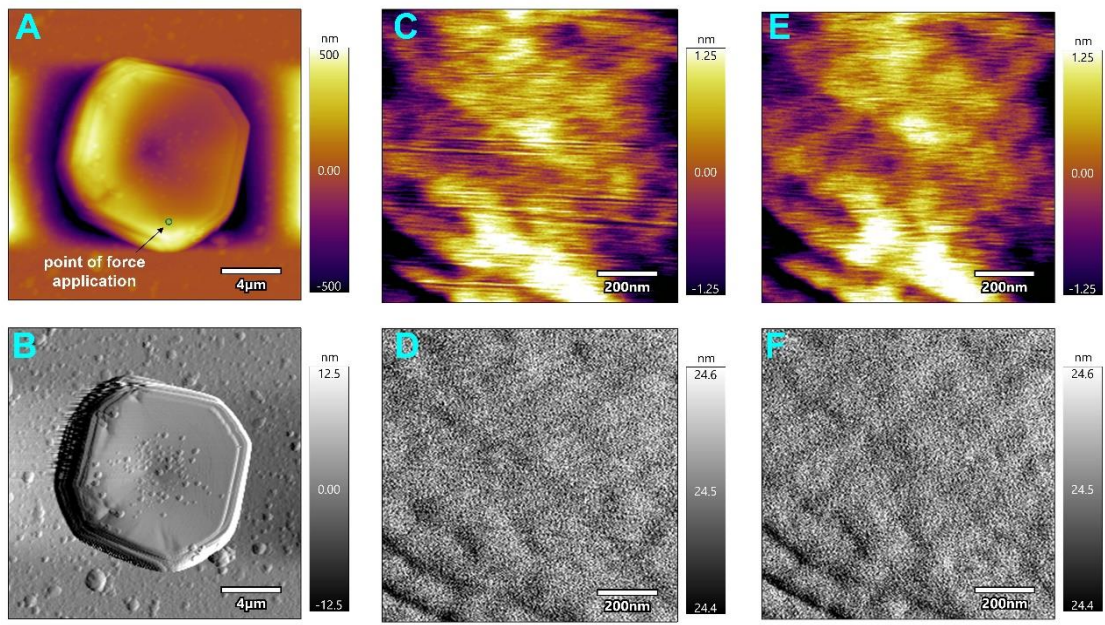

**Figure S3**. AFM measurements of MAPbBr$_3$ single crystal. (A) Height and (B) amplitude of the MAPbBr$_3$ single crystal; (C) morphology and (D) amplitude of the pristine area at the point of force application; (E) morphology and (F) amplitude of the same area after applying a 20 μN force on the tip (150 nm tip diameter).

In order to discard surface damage after applying force, the surface topography of the MAPbBr$_3$ single crystal was measured by atomic force microscopy (AFM) in contact mode, using the same probe as for the ORCA (conductivity) measurement. First, we scanned a 1×1 μm$^2$ area on the surface of MAPbBr$_3$ single crystal, then we picked a point to apply increasing force (0.2 ~ 20 μN) through the diamond tip. Surface topography was scanned after each stress application. It can be seen that, even after applying a force of 20 μN to the AFM, there is no visible indent on the surface of the sample.

## S5. I-V curves of MAPB crystal under AFM tip forces

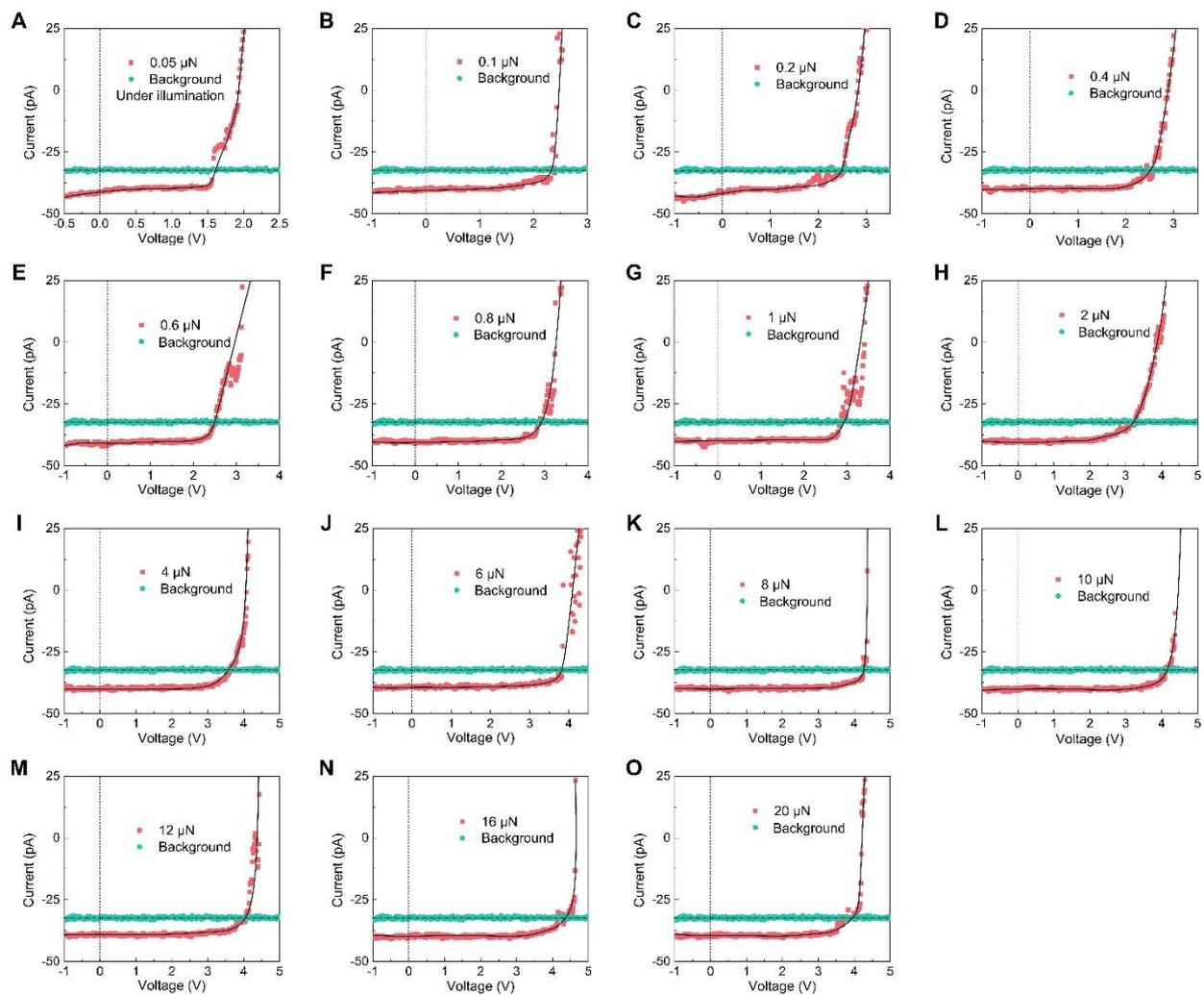

**Figure S4**. The I-V curves of MAPB crystal under different AFM tip forces (150 nm tip diameter), in dark (background) and with illumination.

## S5. Current and voltage *vs* time for given set-points

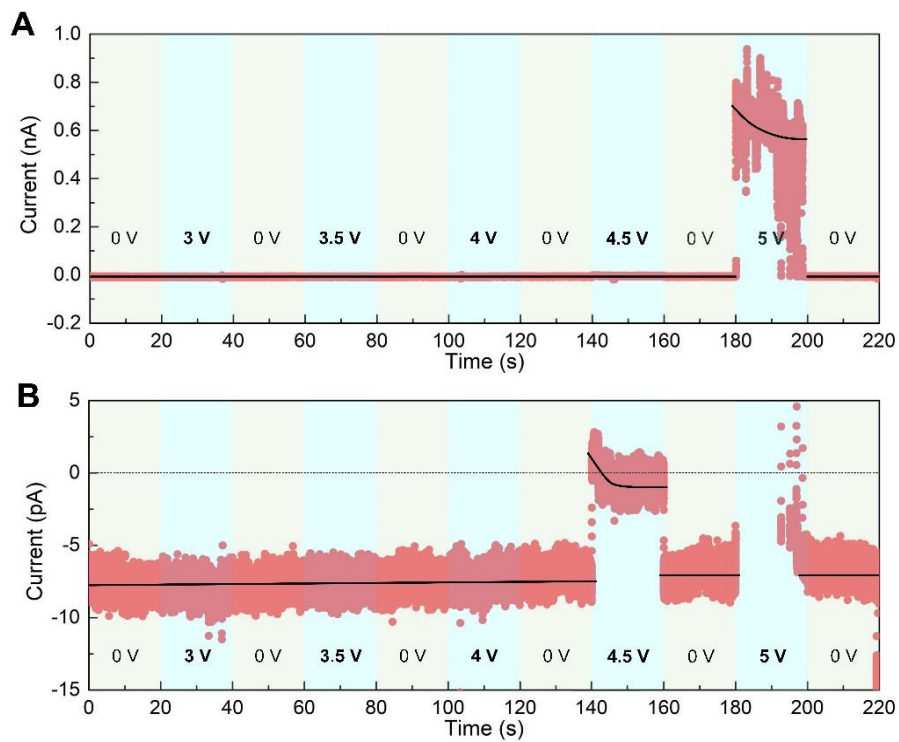

**Figure S5.** (a) Current variations as a function of time at different voltages, for an applied force of 20 μN via an AFM probe of 150 nm tip diameter. (b) The same with a different (more zooomed-in) vertical axis. The results show that the steady-state values do not differ significantly from those measured in the I-V cycles and the above-bandgap photovoltage is not a transient state.

## S6. Flexo-photovoltaic effect of MAPB single crystal by working with a sharper tip

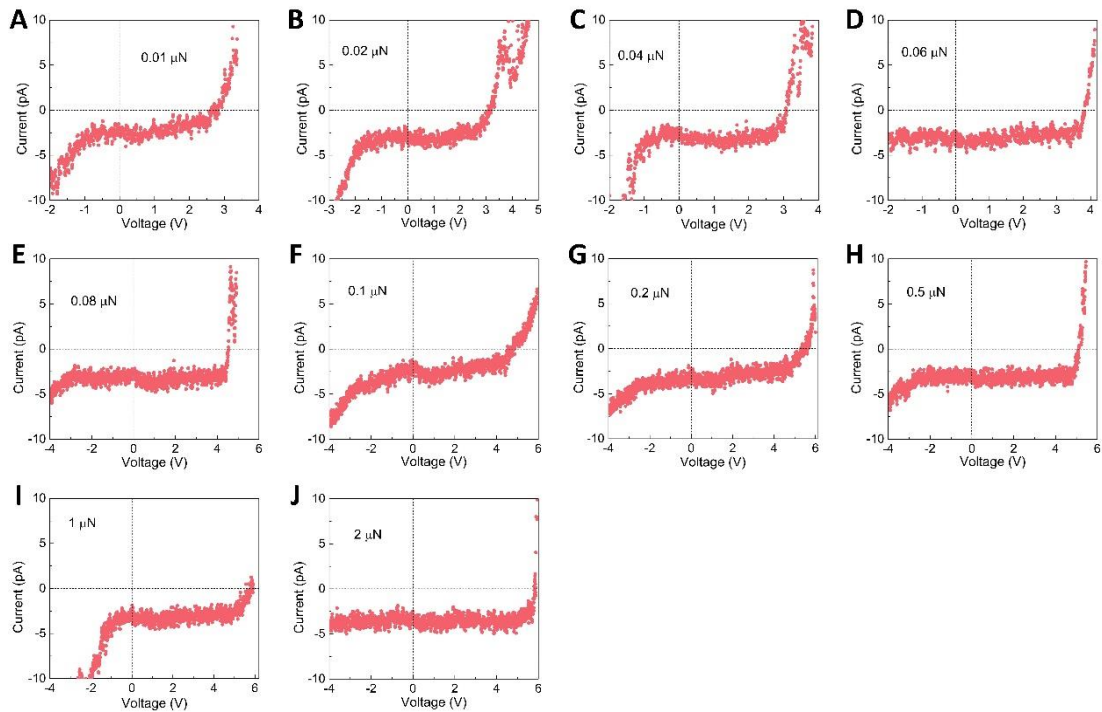

**Figure S6**. The I-V curves of MAPB crystal under different AFM tip forces. The experiment is carried out under illumination. The background current from the AFM has been subtracted. A Ti/Ir coated conductive probe (Model: ASYELEC-01-R2) with a diameter of ~25 nm was used to apply pressure to the MAPB single crystal.

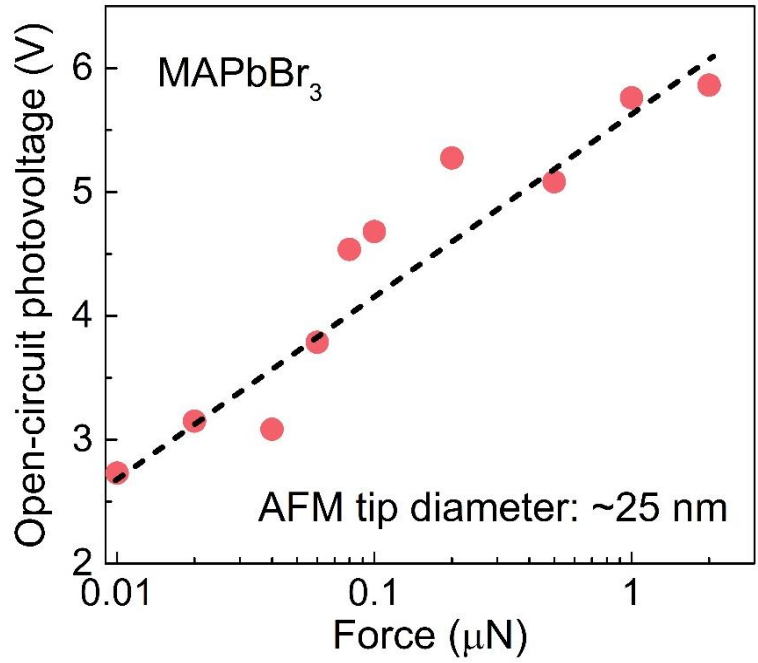

**Figure S7**. Open-circuit photovoltage as a function of indentation force with a sharper (25 nm diameter) AFM probe.

## S7. Finite Element Method (FEM) simulations of contacts between AFM tips and MAPB single crystals.

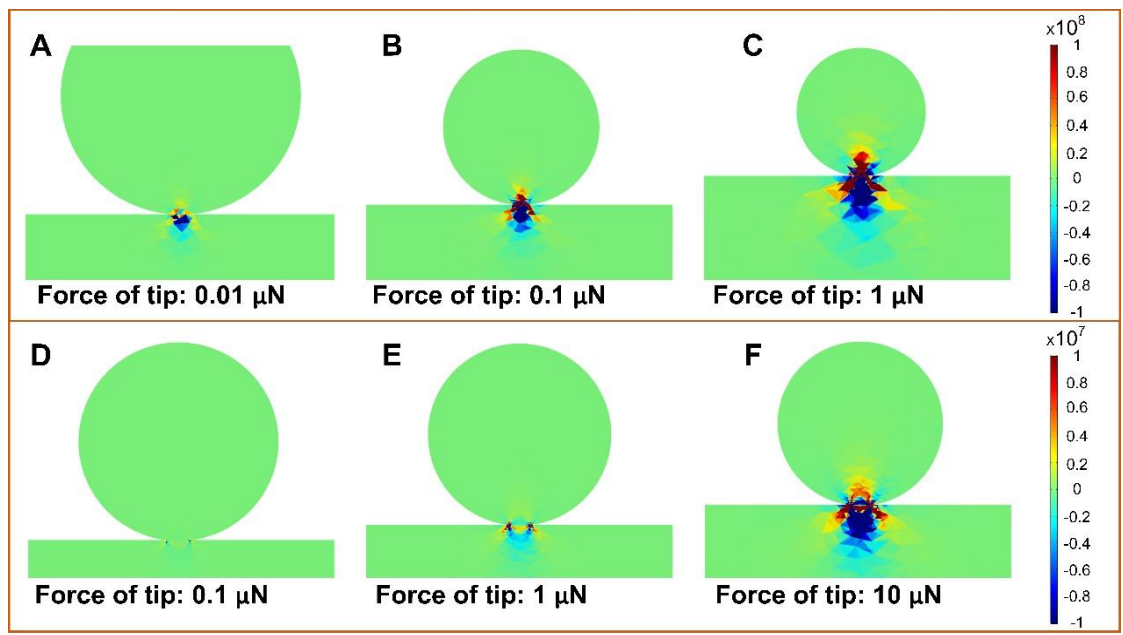

**Figure S8**. FEM analysis of longitudinal strain gradient in the thickness direction for MAPB crystal under the indentation force of two spheres of different diameter, representing AFM tips of different sharpness. (A, B, C) 25 nm diameter tip, 0.01 µN, 0.1 µN and 1 µN forces, respectively. (D, E, F) 150 nm diameter tip, 0.1 µN, 1 µN and 10 µN forces, respectively. The colorbar quantifies, in m$^{-1}$, the derivative of the vertical strain in the vertical direction, $\partial\varepsilon_{zz}/\partial z$.

The deformation of the crystal under different AFM tips was calculated by finite element simulation. Schematic diagrams and FEM calculations are shown in Fig. S8, using spheres of 25nm and 150nm diameter respectively to simulate the two kinds of probe tips we used. Cuboids (width × length × thickness: 200 nm × 200 nm × 150 nm for 25 nm tip, and 400 nm × 400 nm × 300 nm for 150 nm tip) are set up to represent MAPB single crystals. Si (elastic modulu: 200 GPa, poisson's ratio: 0.28, density: 2330 kg/m$^3$) and MAPB (elastic modulu: 32.2 GPa, poisson's ratio: 0.3, [S1] density: 3870 kg/m$^3$) material parameters were used to set the spheres and cuboids respectively. Fig. S8 maps the strain gradient $\partial\varepsilon_{zz}/\partial z$ distribution on the section of YZ under different tip forces, where $\varepsilon_{zz}$ is strain in the direction of thickness. With a sharper tip, a significant increase in strain gradient is obtained even with a smaller force.